\documentclass[print]{mn2e}
\usepackage{epsfig}
\usepackage{color}

\begin{document}
\title[Afterglow Neutrino Emission from LL-GRBs]{Neutrino emission from
a GRB afterglow shock during an inner supernova shock breakout}
\author[Y.W. Yu, Z.G. Dai and X.P. Zheng]{Y.W. Yu$^{1,2}$\thanks{yuyw@nju.edu.cn (YWY)},
Z.G. Dai$^1$\thanks{dzg@nju.edu.cn (ZGD)}, and X.P. Zheng$^2$\thanks{zhxp@phy.ccnu.edu.cn (XPZ)}
\\$^1$Department of Astronomy, Nanjing University, Nanjing 210093,
China
\\$^2$Institute of Astrophysics, Huazhong Normal
University, Wuhan 430079, China}
\maketitle

\begin{abstract}
The observations of a nearby low-luminosity gamma-ray burst (GRB) 060218 associated with supernova SN 2006aj may
imply an interesting astronomical picture where a supernova shock breakout locates behind a relativistic GRB
jet. Based on this picture, we study neutrino emission for early afterglows of GRB 060218-like GRBs, where
neutrinos are expected to be produced from photopion interactions in a GRB blast wave that propagates into a
dense wind. Relativistic protons for the interactions are accelerated by an external shock, while target photons
are basically provided by the incoming thermal emission from the shock breakout and its inverse-Compton
scattered component. Because of a high estimated event rate of low-luminosity GRBs, we would have more
opportunities to detect afterglow neutrinos from a single nearby GRB event of this type by IceCube. Such a
possible detection could provide evidence for the picture described above.
\end{abstract}

\section{introduction}
Gamma-ray burst (GRB) 060218 associated with SN 2006aj discovered by
Swift (Campana et al. 2006) provides a new example of low-luminosity
GRBs (LL-GRBs), as its isotropic equivalent energy
($\sim6\times10^{49}$ erg) is 100 to 1000 times less but its
duration ($T_{90}=2100\pm100$ s) is much longer than those of
conventional high-luminosity GRBs. More interestingly, besides an
usual non-thermal component in its early X-ray spectrum, a
surprising thermal component was observed by the Swift XRT during
both burst and afterglow phases. Fitting with a blackbody spectrum,
the temperature of this thermal component was inferred to be
$kT\sim0.17$ keV during the first 3 ks. When $t>$10 ks, however, the
peak energy of the blackbody decreased and then passed through the
Swift UVOT energy range at $\sim$100 ks (Campana et al. 2006;
Blustin 2007).

To explain the prompt emission, in principle, a model based on the
internal dissipation of relativistic ejecta may be valid in the case
of GRB 060218. The relativistic GRB ejecta, which interacts with a
dense wind surrounding the progenitor, has also been required to
understand a power-law decaying afterglow in X-ray and radio bands
about $\sim$10 ks after the burst, although some complications
beyond the standard afterglow model are involved (Soderberg et al.
2006; Fan et al. 2006; Waxman et al. 2007). Furthermore, as
suggested by Wang \& M\'esz\'aros (2006), the soft X-ray thermal
emission could arise from a shock breakout, namely, a hot cocoon
that breaks out from the supernova ejecta and the stellar wind. In
detail, a more rapid part of a jet moving in the envelope and the
dense wind is accelerated to a highly-relativistic velocity to
produce the GRB, while a slower part of the jet together with the
outermost parts of the envelope becomes a mildly relativistic cocoon
(M\'esz\'aros \& Rees 2001; Ramirez-Ruiz et al. 2002; Zhang et al.
2003), which locates behind the GRB blast wave.

Although the GRB blast wave runs in front of the shock breakout, the
thermal emission from the latter outshines the former persistently
until the emission of the breakout is switched off. Thus, the
emission properties of the GRB blast wave (consisting of external
shock-accelerated electrons and protons) should be influenced by the
incoming thermal photons significantly during both burst and early
afterglow phases. On one hand, the cooling of the relativistic
electrons, which upscatter the thermal photons, could be dominated
by inverse-Compton (IC) radiation rather than synchrotron radiation
(Wang \& M\'esz\'aros 2006). On the other hand, inferred from the
observations, the intensity of the thermal emission could be
comparable in the same band to the one of the prompt emission due to
internal dissipations and much larger than the one of the afterglow
emission due to an external shock. Therefore, the thermal photons as
target photons for photopion interactions could also play an
important role in the energy loss of the relativistic protons and
thus influence or even dominate neutrino emission of the GRB blast
wave.

It has been widely studied that conventional GRBs in the standard
internal-external shock model emit high energy neutrinos during the
burst, early afterglow, and X-ray flare phases (Waxman \& Bahcall
1997, 2000; Dai \& Lu 2001; Dermer 2002; Dermer \& Atoyan 2003;
Asano 2005; Murase \& Nagataki 2006a, 2006b; Murase 2007; Gupta \&
Zhang 2007). In contrast to the conventional GRBs, LL-GRBs may have
a much higher event rate (several hundred to thousand events $\rm
Gpc^{-3}yr^{-1}$), which is inferred from the fact that two typical
nearby LL-GRBs, i.e., GRB 060218 and GRB 980425, have been observed
within a relatively short period of time (Cobb et al. 2006;
Soderberg 2006; Liang et al. 2007). The high event rate implies that
the contribution from LL-GRBs to the diffuse neutrino background may
be important and that we have more opportunities to detect neutrinos
from a very nearby single LL-GRB event. Therefore, Murase et al.
(2006) and Gupta \& Zhang (2006) recently studied the neutrino
emission properties of LL-GRBs during their burst phase using the
internal shock model, in which the target photons for photopion
interactions are mainly provided by internal shock-driven
non-thermal emission. In this paper, however, we will focus on the
early afterglow neutrino emission. During this phase, relativistic
protons are accelerated by an external shock (rather than internal
shocks) and target photons are dominated by the incoming thermal
emission and even its IC scattered component.

This paper is organized as follows. In section 2 we briefly describe
the dynamics of a GRB blast wave propagating into a surrounding
dense wind. In section 3, we give the photon distribution in the
blast wave by considering the incoming thermal emission and its IC
scattered component, but the weak synchrotron radiation of the
electrons is ignored. In section 4, neutrino spectra are derived
formally with an energy loss timescale of protons due to photopion
interactions. In section 5, we calculate the timescale and then the
neutrino spectra using the target photon spectra obtained in section
3 and an experiential fitting formula of the cross section of
photopion interactions. In addition, the peak of a neutrino spectrum
is also estimated analytically by using $\Delta-$approximation.
Finally, a summary is given in section 6.

\section{dynamics of a GRB blast wave}
We consider a GRB jet with isotropic equivalent energy
$E=10^{50}E_{50}~{\rm erg}$ (hereafter $Q_{x}=Q/10^{x}$) expanding
into a dense wind medium with density profile $\rho(r)=Ar^{-2}$.
Here, the coefficient $A$ is determined by the mass loss rate and
velocity of the wind of the progenitor, i.e., $A=\dot{M}/4\pi v_{\rm
w}=5.0\times10^{11}{\rm g~cm^{-1}}A_{*}$, where
$A_{*}\equiv[\dot{M}/(10^{-5}{M_{\odot}~\rm yr^{-1}})][v_{\rm
w}/(10^3\rm km~s^{-1})]^{-1}$. From Dai \& Lu (1998) and Chevalier
\& Li (2000), we get the Lorentz factor and radius of the GRB blast
wave (i.e., the external-shocked wind gas) respectively as
\begin{equation}
\Gamma=\left({9E\over128\pi
Ac^3t}\right)^{1/4}=3.6~E_{50}^{1/4}A_{*}^{-1/4}t_3^{-1/4},
\end{equation}
\begin{equation}
r=\left({9Et\over2\pi Ac}\right)^{1/2}=3.1\times10^{15}{\rm
cm}~E_{50}^{1/2}A_{*}^{-1/2}t_3^{1/2}.
\end{equation}
They satisfy $r=8\Gamma^2ct$, which gives rise to a relationship,
$t'=(16/3)\Gamma t$, between the dynamic time $t'$ measured in the
rest frame of the blast wave and the observed time $t$ (Dai \& Lu
1998).

As the circum-burst wind materials are swept up and shocked, most of the heated electrons before cooling
concentrate at the minimum Lorentz factor $\gamma'_{e, m}\sim\bar{\epsilon}_{e}{m_{p}\over
m_{e}}\Gamma=659~\bar{\epsilon}_{e,-1}E_{50}^{1/4}A_{*}^{-1/4}t_3^{-1/4}$. The symbol
$\bar{\epsilon}_{e}\equiv\epsilon_{e}(p-2)/(p-1)$, where $\epsilon_{e}$ is the usual equipartition factor of the
hot electrons and $p$ is the electron's energy distribution index (where $p>2$ is only considered). Meanwhile, a
fraction $\epsilon_{B}$ of the internal energy is assumed to be occupied by a magnetic field, and then the
strengthen of the magnetic field is calculated by $B'\sim(32\pi \epsilon_{B}\Gamma^2\rho c^2)^{1/2}=78{\rm
G}~\epsilon_{B,-1}^{1/2}E_{50}^{-1/4}A_{*}^{3/4}t_3^{-3/4}$. Finally, the other energy (a fraction of
$\epsilon_p=1-\epsilon_e-\epsilon_B$) is carried by the accelerated protons. For these protons, we can estimate
their maximum energy by $E_{p,\rm max}=2eB'r/3=4.8\times10^{10}{\rm
GeV}~\epsilon_{B,-1}^{1/2}E_{50}^{1/4}A_{*}^{1/4}t_3^{-1/4}$ by equating the acceleration time to the shorter of
the dynamic time and the synchrotron cooling time (Razzaque el al. 2006). However, the minimum energy of the
protons is unknown, but the corresponding Lorentz factor $\gamma'_{p,\rm min}$ is thought to be close to
 $\sim\Gamma$.

\section{photon emission}
The photons in the GRB blast wave have two origins, i.e., the blast
wave self and the inner supernova shock breakout. The electrons in
the blast wave emit photons via synchrotron and IC scattering
processes. Moreover, as analyzed by Wang \& M\'esz\'aros (2006), the
synchrotron radiation (peaking within X-ray band) of the blast wave
electrons is inferred from the observations to be much weaker than
the incoming thermal emission, and thus the cooling of the electrons
should be dominated by their IC scattering off the thermal photons.
Therefore, in following calculations, we consider the thermal
emission and its subsequent IC scattered component only.

The properties of the supernova shock breakout have been unclear to
date. We suppose that it has a constant blackbody temperature of
$kT=0.1 {\rm keV}(kT)_{-1}$ and a constant radius of the emission
region of $R=10^{12}{\rm cm}R_{12}$. The lifetime $t_{\rm SB}$ of
this high-temperature emission is about thousands of seconds, which
is considered to be several to several ten times longer than the
duration of the GRB. Then, the isotropic equivalent luminosity and
energy of the shock breakout can be estimated by $L_{\rm SB}=4\pi
R^2\sigma T^4=1.3\times 10^{45}{\rm erg~s^{-1}}~(kT)_{-1}^4R_{12}^2$
and $E_{\rm SB}=1.3\times 10^{48}{\rm erg}~(kT)_{-1}^4R_{12}^2t_{\rm
SB,3}$, respectively. Meanwhile, it is easy to write the
monochromatic number density of these thermal photons at the
breakout as
\begin{equation}
n(E_{\gamma})={8\pi\over h^3c^3}{E_{\gamma}^2\over
\exp(E_{\gamma}/kT)-1}= {8\pi k^2T^2\over
h^3c^3}\phi\left({E_{\gamma}\over kT}\right),\label{nphbo}
\end{equation}
where the function $\phi(x)={x^2/(e^{x}-1})$. Assuming that the
photons propagate freely before they reach the GRB blast wave at
radius $r$, we can calculate the density of the thermal photons in
the blast wave by multiplying a factor $(R/r)^2$ to Eq.
(\ref{nphbo}). Subsequently, after Lorentz transformation, we obtain
the density of the incoming photons in the blast wave measured in
its rest frame by
\begin{equation}
n'_{\rm in}(E'_{\gamma,\rm in})={R^2\over r^2}n(\Gamma
E'_{\gamma,\rm in})={R^2\over r^2}{8\pi k^2T^2\over
h^3c^3}\phi\left({3E'_{\gamma,\rm in}\over E'_{\gamma,\rm
pk1}}\right),\label{nphin}
\end{equation}
where $E'_{\gamma,\rm pk1}\equiv 3kT/\Gamma$ is the peak energy of
the black body spectrum. When these photons cross the blast wave, a
part of them should be upscattered by the relativistic electrons.
The energy of the IC scattered photons can be estimated by
${E'}_{\gamma, \rm IC}=2{\gamma'}_{e,m}^2E'_{\gamma,\rm in}$ and the
corresponding density by
\begin{eqnarray}
n'_{\rm IC}({E'}_{\gamma, \rm IC}) & = & {\tau\over
2{\gamma'}_{e,m}^2} n'_{\rm in}\left({{E'}_{\gamma, \rm IC}}\over
2{\gamma'}_{e,m}^2\right)\nonumber \\ & = & {\tau\over
2{\gamma'}_{e,m}^2}{R^2\over r^2}{8\pi k^2T^2\over
h^3c^3}\phi\left({3E'_{\gamma,\rm IC}\over E'_{\gamma,\rm
pk2}}\right),\label{nphic}
\end{eqnarray}
where $E'_{\gamma,\rm pk2}\equiv 6{\gamma'}_{e,m}^2kT/\Gamma$. The
probability of the scattering is represented by the photon optical
depth of the blast wave, $ \tau=\sigma_{\rm
T}({A/m_pr})=6.5\times10^{-5}~E_{50}^{-1/2}A_{*}^{3/2}t_3^{-1/2}$,
where $\sigma_{\rm T}$ is the Thomson cross section. According to
this estimation, Wang \& M\'esz\'aros (2006) predicted that the
early afterglow spectra of GRB 060218-like GRBs may have a bimodal
profile peaking at
\begin{equation}
{E}_{\gamma,\rm pk1}=0.3~{\rm keV}~(kT)_{-1}\label{egb1}
\end{equation}
and
\begin{equation}
{E}_{\gamma,\rm pk2}=0.26~{\rm GeV}~\bar{\epsilon}_{\rm
e,-1}^{2}(kT)_{-1}E_{50}^{1/2}A_{*}^{-1/2}t_3^{-1/2}.\label{egb2}
\end{equation}
Thus, a significant sub-GeV or GeV emission component accompanying
the thermal emission would be detectable with the upcoming
\textit{Gamma-ray Large Area Space Telescope}, which could provide
evidence for the GRB jet.

\section{neutrino production}
Since relativistic protons in the GRB blast wave are immersed in the photon field described above, the protons
would lose their energy to produce mesons such as $\pi^0$ and $\pi^\pm$ etc, and subsequently generate neutrinos
by the decay of $\pi^{\pm}$, i.e., $\pi^{\pm}\rightarrow \mu^{\pm}+\nu_{\mu}(\bar{\nu}_{\mu})\rightarrow
e^{\pm}+\nu_{e}(\bar{\nu}_{e})+\bar{\nu}_{\mu}+\nu_{\mu}$. During these processes, the energy loss rate of a
proton with energy $E'_{p}=\gamma'_{p}m_{p}c^2$ can be calculated by (Waxman \& Bahcall 1997)\footnote{To obtain
this expression, an isotropic target photon field is required. However, in our model, the radially incoming
photon field is seen by the protons in the blast wave anisotropically. This gives an extra complication for a
more realistic consideration. For simplicity, we ignore the anisotropic effect in our calculations.}
\begin{eqnarray}
{t'}_{\pi}^{-1} & \equiv & -{1\over E'_{p}}{dE'_{p}\over dt'}\nonumber \\
& = & {c\over 2{\gamma'}_{p}^2}\int_{\tilde{E}_{\rm
th}}^{\infty}\sigma_{\pi}(\tilde{E})\xi(\tilde{E})\tilde{E}\nonumber
\\& & \times \left[\int_{\tilde{E}/2\gamma'_{p}}^{\infty}n'(E'_{\gamma})
{E'}_{\gamma}^{-2}dE'_{\gamma}\right]d\tilde{E},\label{tpg1}
\end{eqnarray}
where $\sigma_{\pi}(\tilde{E})$ is the cross section of photopion
interactions for a target photon with energy $\tilde{E}$ in the
proton's rest frame, $\xi$ is the inelasticity defined as the
fraction of energy loss of a proton to the resultant pions, and
$\tilde{E}_{\rm th}=0.15\rm GeV$ is the threshold energy of the
interactions. Equation (\ref{tpg1}) yields that the energy of the
protons decreases as
$\exp\left[-\int_0^{t'}({dt'/{t'}_{\pi}})\right]$. In our scenario,
if the shock-breakout emission could last for a period of $t_{\rm
SB}$, the fraction of the energy loss of the protons to pions could
be calculated by
\begin{equation}
f_{\pi}=1-\exp\left(-\int_0^{t'_{\rm
SB}}{dt'\over{t'}_{\pi}}\right),\label{fpi1}
\end{equation}
where $t'_{\rm SB}=(16/3)\Gamma t_{\rm SB}$. In order to calculate
$t'_{\pi}$, the crucial input in the model is the target photon
spectrum $n'(E'_{\gamma})$. From Eqs. (\ref{nphin}) and
(\ref{nphic}), we know that $n'(E'_{\gamma})$ depends on both $r$
and $\Gamma$ and thus the value of $t'_{\pi}$ could evolve with
time. However, if $t'_{\pi}$ is independent of time or varies with
time slowly, Eq. (\ref{fpi1}) can be also approximated by
\begin{equation}
f_{\pi}\approx 1-\exp(-t'_{\rm SB}/{t'}_{\pi})\approx \min[t'_{\rm
SB}/{t'}_{\pi},1]\label{fpiapp}
\end{equation}
as usual, especially for analytical calculations.

To be specific, the energy loss of the protons is shared by
$\pi^{\pm}$ and $\pi^{0}$ with a certain ratio. Unfortunately, it is
not easy to fix this ratio due to the complications arising from
various single-pion and multipion production processes. In following
calculations, we simply take it to be a constant,
$\pi^{\pm}:\pi^0=2:1$, as in Asano (2005). Furthermore, two
resultant muon-neutrinos from the decay of a $\pi^{\pm}$ could
inherit half of the pion's energy roughly evenly. Therefore, we can
relate the neutrino energy $E_{\nu}$ to the energy loss of the
primary proton by
\begin{equation}
E_{\nu}={1\over4}\xi E_{p},\label{enu}
\end{equation}
and give an observed time-integrated muon-neutrino spectrum by
\begin{equation}
E_{\nu}^2\phi_{\nu}\equiv  {1\over 4\pi
D_{l}^2}E_{\nu}^2{dN_{\nu}\over dE_{\nu}}={1\over 4\pi
D_{l}^2}{f_{\pi}\over 3} {E}_{p}^2{dN_{p}\over
dE_{p}}\label{nuspectra},
\end{equation}
where $D_l$ is the luminosity distance of the burst. As usual, we
assume the energy distribution of the shock-accelerated protons to
be $({dN_{p}/ dE_{p}})\propto{E}_{p}^{-2}$, where the proportional
coefficient can be calculated by $\epsilon_pE/\ln(E_{p,\rm
max}/E_{p,\rm min})$.

In addition, because of the presence of the magnetic field, the
ultrahigh energy pions and muons would lose their energy via
synchrotron radiation before decay. This leads to breaks in the
neutrino spectrum at (Murase 2007)
\begin{eqnarray}
E_{\nu, \rm b}^{s\pi} & = & {1\over4}E_{\rm
\pi,b}={1\over4}\Gamma\left({6\pi m_{\pi}^5c^5\over \sigma_{\rm
T}m_{\rm e}^2B'^2\tau_{\pi}}\right)^{1/2}\nonumber \\ & = &
1.2\times 10^{9}{\rm
GeV}~\epsilon_{B,-1}^{-1/2}E_{50}^{1/2}A_{*}^{-1}t_3^{1/2},
\end{eqnarray}
\begin{eqnarray}
E_{\nu, \rm b}^{s\mu} & = & {1\over3}E_{\mu, \rm
b}={1\over3}\Gamma\left({6\pi m_{\mu}^5c^5\over \sigma_{\rm T}m_{\rm
e}^2B'^2\tau_{\mu}}\right)^{1/2}\nonumber \\ & = & 8.9\times
10^{7}{\rm
GeV}~\epsilon_{B,-1}^{-1/2}E_{50}^{1/2}A_{*}^{-1}t_3^{1/2},
\end{eqnarray}
where $\tau_{\pi}=2.6\times 10^{-8}$s and $\tau_{\mu}=2.2\times 10^{-6}$s are the mean lifetimes of pions and
muons in their rest frames. Above $E_{\nu, \rm b}^s$, the neutrino flux would be suppressed by a factor
$(E_{\nu}/E_{\nu, \rm b}^s)^{-2}$ (Rachen \& M\'esz\'aros 1998; Razzaque et al. 2006). However, as pointed out
by Asano \& Nagataki (2006), neutral kaons can survive in the magnetic field,  while the ultrahigh-energy
charged pions and muons cool rapidly. Moreover, because kaons have a larger rest mass than pions and muons,
charged kaons can reach higher energy although they also suffer from synchrotron cooling. Thus, decay of kaons,
which is not taken into account in our calculations, may dominate neutrino emission above $\sim10^8-10^9$GeV.

Now, by inserting Eqs. (\ref{nphin}) and (\ref{nphic}) into Eq.
(\ref{tpg1}) and then into Eq. (\ref{fpi1}) to get $f_\pi$, we can
easily obtain the observed neutrino spectra from Eq.
(\ref{nuspectra}) for our scenario. The remaining task is only to
express the cross section $\sigma_{\pi}(\tilde{E})$ and inelasticity
for photopion interactions.

\section{results}
Since the cross section of photopion interactions peaks at
$\tilde{E}_{\Delta}\simeq0.3\rm GeV$ due to the
$\Delta$(1232)-resonance, the integration over $\tilde{E}$ in Eq.
(\ref{tpg1}) can be roughly approximated by
\begin{equation}
{t'}_{\pi}^{-1}\approx{c\over 2{\gamma'}_{p}^2}\sigma_{\pi,\Delta}\xi_{
\Delta}\tilde{E}_{\Delta}\delta\tilde{E}\int_{\tilde{E}_{\Delta}/2\gamma'_{
p}}^{\infty}n'(E'_{\gamma}){E'}_{\gamma}^{-2}dE'_{\gamma},\label{tpg}
\end{equation}
where $\sigma_{\pi,\Delta}\approx0.5\rm mbarn$,
$\xi_{\Delta}\approx0.2$, and the peak width is about
$\delta\tilde{E}\approx0.2\rm GeV$. Inserting Eq. (\ref{nphin}) or
(\ref{nphic}) into Eq. (\ref{tpg}), we use the approximative formula
$f_{\pi}=\min[t'_{\rm SB}/t'_{\pi}$,1] to obtain
\begin{eqnarray}
{f}_{\pi} &=& \min\left\{{16\over3}\varsigma{R^2\over r^2}{8\pi
k^2T^2\over h^3c^2}{2E'_{\gamma,\rm pk}\over
3\tilde{E}_{\Delta}}\sigma_{\pi,\Delta}\xi_{\Delta}\delta\tilde{E}\Gamma
t_{\rm SB}\right. \nonumber\\
&&\left.\times\varepsilon_{*}^2\left[\varepsilon_{*}-\ln
(e^{\varepsilon_{*}}-1)\right],~1\right\}
.\label{fpi}
\end{eqnarray}
where $\varsigma=1$ and $\tau/(2{\gamma'}_{e,m}^2)$ for pre- and
post-upscattered target photons, respectively. The dimensionless
variable $\varepsilon_{*}$ is defined by
$\varepsilon_{*}\equiv3\tilde{E}_{\Delta}/(2\gamma'_{p}E'_{\gamma,\rm
pk}) = {3\xi_{\Delta}\Gamma^2\tilde{E}_{\Delta}m_{p}c^2
/(8{E}_{\nu}E_{\gamma,\rm
 pk}})$. In the case of $f_{\pi}<1$,
 the peak value of $f_{\pi}$ reading
\begin{equation}
f_{\pi,\rm pk}=3\varsigma{R^2\over r^2}{8\pi k^2T^2\over h^3c^2}{2E'_{\gamma,\rm pk}\over
3\tilde{E}_{\Delta}}\sigma_{\pi,\Delta}\xi_{\Delta}\delta\tilde{E}\Gamma t_{\rm SB}
\end{equation}
is at $\varepsilon_*=1.8$, which gives rise to the relationship between the peak energies of the neutrino and
photon spectra as $E_{\nu,\rm pk}E_{\gamma,\rm pk}=0.01\Gamma^2\rm GeV^2$. Considering the bimodal distribution
of the target photons peaking at $E_{\gamma,\rm pk1}$ and $E_{\gamma,\rm pk2}$, two peaks are also expected in
the resultant neutrino spectrum but only the one determined by $E_{\gamma,\rm pk1}$ could fall into the high
energy range ($E_{\nu}>\rm TeV$) of our interest at
\begin{equation}
{E}_{\nu,\rm pk}=4.9\times10^5{\rm
GeV}~(kT)_{-1}^{-1}E_{50}^{1/2}A_{*}^{-1/2}t_{\rm SB,3}^{-1/2}.
\end{equation}
In other words, the target photons for photopion interactions of
interest are contributed by the incoming thermal emission mainly.
The value of the differential neutrino fluence at ${E}_{\nu,\rm pk}$
reads
\begin{equation}
[E_{\nu}^2\phi_{\nu}]_{\rm pk}=2.0\times 10^{-6}{\rm
erg~cm^{-2}}~\epsilon_p(kT)_{-1}^3R_{12}^2A_{*}D_{l,25.5}^{-2},
\end{equation}
which is calculated by using the peak value of $f_{\pi}$ as
\begin{equation}
f_{\pi,\rm pk}=0.02~(kT)_{-1}^3R_{12}^{2}E_{50}^{-1}A_{*}.\label{fpipk}
\end{equation}
On the other hand, when $f_{\pi}=1$, $E_{\nu}^2\phi_{\nu}$ would
reach an upper limit as $1.2\times10^{-4}{\rm
erg~cm^{-2}}~\epsilon_pE_{50}D_{l,25.5}^{-2}$, which is determined
by the total energy carried by the protons in the GRB blast wave.

Although it is convenient and effective to use the $\Delta-$approximation to estimate the peak of a neutrino
spectrum, the $\Delta-$approximation would lead to an remarkable underestimation of the neutrino flux above the
peak energy due to the non-zero cross section of photopion interactions in high energy regions. So, for more
careful calculations, we provide an experiential fitting formula for the cross section as shown in Eq.
(\ref{sigtotal}), which is extrapolated from experimental data taken from particle data group (Yao et al. 2006).
However, since we can not find a simple expression for the inelasticity, we take $\xi =0.2$ for all energy
regions roughly, which may leads to a mild underestimation of the neutrino flux in the high energy regions.
Finally, with these inputs, we plot the observed time-integrated muon-neutrino spectra in Fig. 1. Obviously, two
plateaus exist in the neutrino spectra. To be specific, as shown in the upper panel of Fig.1, the high-energy
plateau is produced by the lower energy thermal photons, while the low-energy plateau is produced by the higher
energy IC scattered photons. In addition, from a comparison shown in the lower panel of Fig. 1, we can see that
the approximation for $f_{\pi}$ in Eq. (\ref{fpiapp}) is feasible to some extent for the thermal seed
photon-dominated photopion interactions, but not for the IC scattered photon-dominated interactions. This
difference of these two kinds of interaction arises from different temporal behaviors of $t'_{\pi}$.

Next let's discuss the detectability of the afterglow neutrinos,
using the following fitting formula for the probability of detecting
muon-neutrinos
by IceCube (Ioka et al. 2005; Razzaque et al. 2004)
\begin{equation}
P_{\nu}=7\times10^{-5}\left({E_{\nu}\over 3.2\times10^4\rm
GeV}\right)^{\beta},\label{deprob}
\end{equation}
where $\beta=1.35$ for $E_{\nu}<3.2\times10^4\rm GeV$, while
$\beta=0.55$ for $E_{\nu}\geq3.2\times10^4\rm GeV$. The number of
muon events from muon-neutrinos above TeV energy is given by
\begin{equation}
N_{\mu}=A_{\rm det}\int_{\rm TeV}\phi_{\nu}P_{\nu}dE_{\nu},
\end{equation}
where $A_{\rm det}\sim1\rm km^2$ is the geometrical detector area.
Inserting Eqs. (\ref{nuspectra}) and (\ref{deprob}) into the above
integral, we obtain $N_{\mu}\sim0.1$ for the parameter set
($E_{50}=1$, $A_{*}=10$, $(kT)_{-1}=2$, $R_{12}=1$ and $t_{\rm
SB,3}=3$) inferred from GRB 060218 for a very nearby LL-GRB at 50
Mpc, where a LL-GRB event is expected to be observed within a
many-years observation. According to this estimation, we expect
optimistically that IceCube may be able to detect afterglow
neutrinos from one LL-GRB event in the following decades. If such a
detection comes true, the afterglow neutrino emission accompanying
the soft X-ray thermal and sub-GeV or GeV emissions from a GRB
060218-like GRB event would provide strong evidence for the picture
that a supernova shock breakout locates behind a relativistic GRB
jet, and further would be used to constrain the model parameters
severely.

Besides the possible detection of neutrinos from a single LL-GRB
event, the contribution to the neutrino background from  LL-GRBs is
also expected to be important. We can estimate the diffuse
muon-neutrino flux arising from afterglow neutrino emission of
LL-GRBs by (Waxman \& Bahcall 1998; Murase et al. 2006)
\begin{eqnarray}
E_{\nu}^2\Phi_{\nu} & \sim & {c\over 4\pi
H_0}{f_{\pi}\over3}f_b\epsilon_pE_p^2{dN_p\over dE_p}R_{\rm
LL}(0)f_{z}\nonumber \\
& = & 2.5\times 10^{-11}{\rm
GeV~cm^{-2}s^{-1}sr^{-1}}~\epsilon_p(kT)_{-1}^3R_{12}^2A_{*}
\nonumber
\\ & & \times f_b\left({R_{\rm LL}(0)\over 500{\rm
Gpc^{-3}yr^{-1}}}\right)\left({f_z\over 3}\right),\label{diffuse}
\end{eqnarray}
where $H_0=71 \rm km~s^{-1}Mpc^{-1}$, $f_b$ is the beaming factor, and $f_z$ is the correction factor for the
possible contribution from high-redshift sources. In the above estimation, the approximative value of $f_{\pi}$
in Eq. (\ref{fpipk}) is applied. By comparing Eq. (\ref{diffuse}) to Eq. (3) of Murase et al. (2006), we find
that, for LL-GRBs, the contribution to the diffuse neutrino background by the early afterglow neutrino emission
may be relatively smaller than or even comparable to (e.g., for model parameters $\epsilon_p=0.6$,
$(kT)_{-1}=2$, $R_{12}=1$, and $A_{*}=10$) that by the burst neutrino emission.
\begin{figure}
\resizebox{\hsize}{!}{\includegraphics{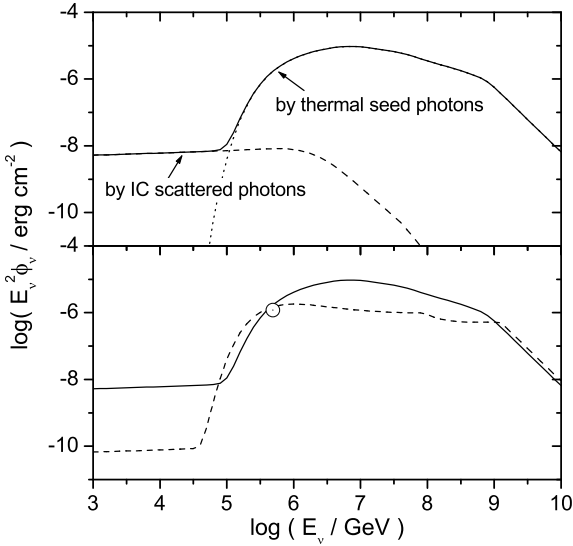}} \caption{The
time-integrated afterglow muon-neutrino
($\nu_{\mu}+\bar{\nu}_{\mu}$) spectra for one GRB event. The solid
lines are calculated by using the expressions for $f_{\pi}$ in Eq.
(\ref{fpi1}) and for $\sigma_{\pi}(\tilde{E})$ in Eq.
(\ref{sigtotal}). \textit{Upper panel}: The contributions to the
total neutrino emission by the two target photon components are
represented by the dashed and dotted lines, respectively.
\textit{Lower panel}: The dashed line is obtained by an
approximation for the time integration as in Eq. (\ref{fpiapp}), and
the peak estimated by the $\Delta-$approximation is labeled by an
open circle. In all cases, we take the model parameters $E_{50}$,
$A_{*}$, $(kT)_{-1}$, $R_{12}$, $D_{l,25.5}$ and $t_{\rm SB,3}$ to
be unity and the equipartition factors ${\epsilon}_{e}=0.3,
\epsilon_{B}=0.1$ and thus $\epsilon_{p}=0.6$.}
\end{figure}

Finally, we would like to refer the reader to neutrino oscillation, which will change neutrino flavor ratio from
$\nu_e : \nu_{\mu} : \nu_{\tau} \simeq 1 : 2 : 0$ at the source to $1 : 1 : 1$ at the earth. This thus leads to
the fact that the observed muon-neutrino fluxes estimated above should be reduced further by a factor of
$\sim2$.

\section{Summary}
The surprising soft X-ray thermal emission during both burst and afterglow phases of GRB 060218/SN 2006aj was
proposed to be due to the breakout from a strong stellar wind of a radiation-dominated shock. This shock
breakout was further thought to locate behind a relativistic GRB jet, which is required by understanding the
burst emission and the power-law decaying afterglow emission. Wang \& M\'esz\'aros (2006) suggested that a
sub-GeV or GeV emission produced by IC scattering of the thermal photons by the relativistic electrons in the
GRB blast wave could give evidence for this astronomical picture. In this paper, we studied another possible
implication, namely, afterglow neutrino emission. The neutrinos are produced by photopion interactions of
relativistic protons, which could be accelerated by a relativistic external shock. The target photons in the
interactions are contributed by the incoming thermal emission and its upscattered component. By considering the
high event rate of LL-GRBs, we argue optimistically that the afterglow neutrinos from very nearby (several tens
of Mpc) LL-GRBs may be detected by IceCube in the following decades. We believe the detection of these expected
afterglow neutrinos is helpful to uncover the nature of GRB 060218-like GRBs.

\section*{Acknowledgements}
We would like to thank the referee for helpful comments and
suggestions. This work is supported by the National Natural Science
Foundation of China (grants 10221001 and 10640420144) and the
National Basic Research Program of China (973 program) No.
2007CB815404. Y.W.Y. is also supported by the Visiting PhD Candidate
Foundation of Nanjing University and partly by the National Natural
Science Foundation of China (grants 10603002 and 10773004).

\appendix
\section{cross section fits}
The fits to the total cross section for $p\gamma$ interactions have
been widely studied (e.g., Rachen 1996; M\"{u}cke et al. 2000). In
physics, this cross section is contributed by resonant excitations
and direct (non-resonant) single-pion production processes in the
resonant energy regions and by statistical multipion production
processes mainly and diffractive scattering slightly in the high
energy region ($\tilde{E}_0>$0.727 GeV).

It is known that the cross section for a resonance is given by the
Breit-Wigner formula (M\"{u}cke et al. 2000)
\begin{equation}
\sigma_R(\tilde{E})={s\over
\tilde{E}^2}{\sigma_0\bar{\Gamma}^2s\over(s-m_{R}^2c^4)^2+\bar{\Gamma}^2s},
\end{equation}
where $\sqrt{s}=(m_p^2c^4+2m_pc^2\tilde{E})^{1/2}$ is the total
energy of the colliding photon and proton in the mass-center frame,
and $m_{R}$ and $\bar{\Gamma}$ are the nominal mass and width of the
resonance, respectively. The coefficient $\sigma_0$ is determined by
the resonance angular momentum and the electromagnetic excitation
strength. For nine important resonances in $p\gamma$ interactions,
we take the related parameters from M\"{u}cke et al. (2000) and list
them in Table 1. Then, the total cross section contributed by these
resonances can be written as
\begin{equation}
\sigma_1(\tilde{E})=\left[1-\exp\left({0.15-x\over0.08}\right)\right]\sum
\sigma_R(\tilde{E}),\label{sig1}
\end{equation}
where the suppression factor in the square bracket represents the
threshold $\tilde{E}_{\rm th}=0.15$GeV of $p\gamma$ interactions
with $x=\tilde{E}/\rm GeV$. Moreover, Rachen (1996) found that the
cross section in the high energy region ($\tilde{E}_0>$0.727 GeV)
can be fitted by\footnote{The coefficients here are mildly different
from Rachen (1996).}
\begin{eqnarray}
\sigma_2(\tilde{E})&=&\left[1-\exp\left({0.727-x\over0.8}\right)\right]
\times\nonumber\\ &&\left(0.067y^{0.081} +0.125y^{-0.453}\right)
~\rm mbarn.\label{sig2}
\end{eqnarray}
where $y=s/\rm GeV^2$. Subtracting the two components expressed by
Eqs. \ref{sig1} and \ref{sig2} from the experimental data taken from
particle data group (Yao et al. 2006), we find the residuals to the
total cross section exhibit a broken power-low behavior, which
yields
\begin{eqnarray}
\sigma_3(\tilde{E})&=&\left[1-\exp\left({0.15-x\over0.08}\right)\right]
\times\nonumber\\ &&0.072x^{-1.6}\left[\left({0.62\over
x}\right)^{20} +
        1\right]^{-0.21}
\times\\
&& \left[\left({0.46\over x}\right)^{20} +
1\right]^{0.225}\left[\left({0.28\over x}\right)^{20} +
1\right]^{-0.45}~\rm mbarn\nonumber.
\end{eqnarray}
Roughly speaking, the above formula could be related with the direct
single-pion production processes. Finally, combining the three
components, we can express the total cross section of $p\gamma$
interactions by
\begin{equation}
\sigma_{\pi}(\tilde{E})=\left\{
\begin{array}{ll}
\sigma_1+\sigma_3,~{\rm for~} 0.15{\rm GeV}<\tilde{E}_0<0.727{\rm GeV};\\
\sigma_1+\sigma_2+\sigma_3,~{\rm for~} \tilde{E}_0>0.727{\rm GeV}.\\
\end{array}
\right.\label{sigtotal}
\end{equation}
We confront this experiential formula with the experimental data in
Fig. A1. It can be seen the fits is good, although the formula can
not describe the detailed physics.
\begin{table}
  \caption[]{Parameters for resonances. }
  \begin{center}\begin{tabular}{cccc}
  \hline
Name&$m_Rc^2$/GeV& $\bar{\Gamma}$/GeV & $\sigma_0/\mu$barn \\
  \hline
$\Delta(1232)$  & 1.231 & 0.11 & 31.125          \\
$N(1440)$       & 1.44  & 0.35 & 1.389          \\
$N(1520)$       & 1.515 & 0.11 & 25.567          \\
$N(1535)$       & 1.525 & 0.1  & 6.948          \\
$N(1650)$       & 1.675 & 0.16 & 2.779          \\
$N(1680)$       & 1.68  & 0.125& 17.508          \\
$\Delta(1700)$  & 1.69  & 0.29 & 11.116          \\
$\Delta(1905)$  & 1.895 & 0.35 & 1.667          \\
$\Delta(1950)$  & 1.95  & 0.3  & 11.116          \\
\hline
  \end{tabular}\end{center}
\end{table}
\begin{figure}
\resizebox{\hsize}{!}{\includegraphics{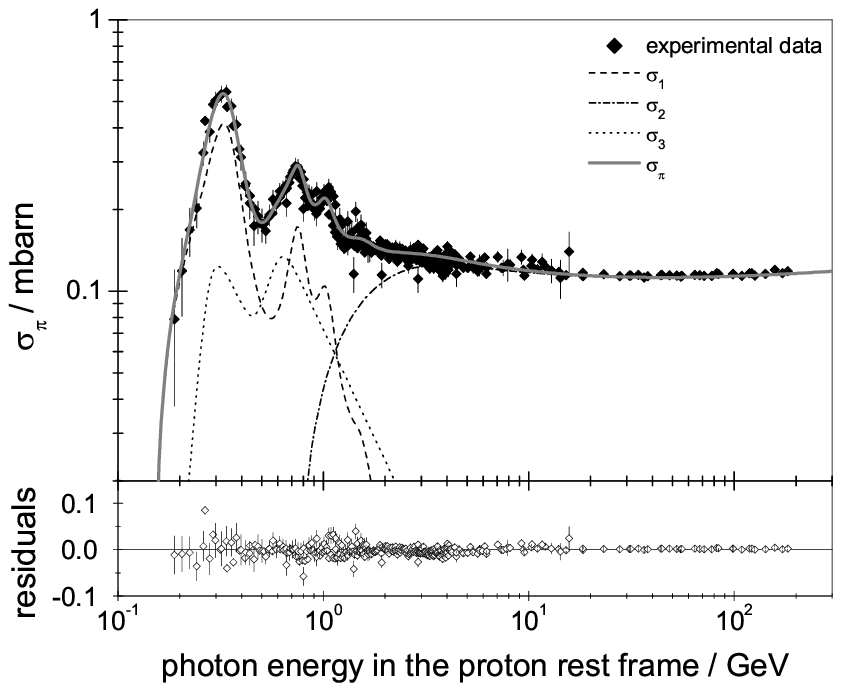}} \caption{Fits for
the total cross section for $p\gamma$ interactions.}
\end{figure}

\end{document}